# Three-dimensional Multiscale Model of Deformable Platelets Adhesion to Vessel Wall in Blood Flow

Ziheng Wu*, Zhiliang Xu*, Oleg Kim*, and Mark Alber *, **

* Department of Applied and Computational Mathematics and Statistics,

University of Notre Dame, Notre Dame, IN 46556, USA

** Department of Medicine, Indiana University School of Medicine, IN 46202, USA

March 23, 2014

Running title: Deformable Platelets Adhesion to Vessel Wall

Address for Correspondence and Reprint Requests:

Mark Alber

Department of Applied and Computational Mathematics and Statistics

University of Notre Dame, Notre Dame, IN 46556, USA

Phone: (574) 631-8371, Email: malber@nd.edu

**Abstract**

When a blood vessel ruptures or gets inflamed, the human body responds by rapidly forming a clot to restrict the loss of blood. Platelets aggregation at the injury site of the blood vessel occurring via platelet-platelet adhesion, tethering and rolling on the injured endothelium is a critical initial step in blood clot formation. A novel three-dimensional multiscale model is introduced and used in this paper to simulate receptor-mediated adhesion of deformable platelets at the site of vascular injury under different shear rates of blood flow. The novelty of the model is based on a new approach of coupling submodels at three biological scales crucial for the early clot formation: novel hybrid cell membrane submodel to represent physiological elastic properties of a platelet, stochastic receptor-ligand binding submodel to describe cell adhesion kinetics and Lattice Boltzmann submodel for simulating blood flow. The model implementation on the GPUs cluster significantly improved simulation performance. Predictive model simulations revealed that platelet deformation, interactions between platelets in the vicinity of the vessel wall as well as the number of functional GPIbα platelet receptors played significant roles in the platelet adhesion to the injury site. Variation of the number of functional GPIbα platelet receptors as well as changes of platelet stiffness can represent effects of specific drugs reducing or enhancing platelet activity. Therefore, predictive simulations can improve the search for new drug targets and help to make treatment of thrombosis patient specific.

**Keywords:** three-dimensional model, platelet adhesion, cell flow interaction, Lattice Boltzmann, stochastic receptor-ligand model, thrombus, blood clot

## 1. Introduction

When a blood vessel ruptures or gets inflamed, the human body responds by rapidly forming a clot to restrict the loss of blood. Blood clots (thrombi) formation is a complex biological process involving an extensive system of biochemical (coagulation) reactions, platelet hydrodynamics, platelet-platelet and platelet-blood vessel wall interactions leading to ligand-receptor adhesion bond formation and platelet activation. Platelet adhesion to the vessel wall is one of the first events associated with formation of hemostatic clots and pathological thrombi.

In this paper, a new 3D multiscale model of platelet-blood flow-vessel wall interactions combining submnodels at three biological scales crucial for the early platelet aggregation is introduced and calibrated to investigate how platelet stiffness, GPIb receptor expression and platelet-platelet interaction affect platelet-wall adhesion quantified in terms of platelet pause time. We implemented a novel approach of combining a recently developed platelet hybrid membrane submodel, the SCE representation of the cytoskeleton network and a continuum description of the lipid bilayer to study the very first step of blood clot formation, the rapid formation of unstable bonds which slow platelets and cause platelet flipping and adhesion to the damaged surface. The hybrid platelet model was also coupled with the Lattice Boltzmann model (LBM) of blood using the immersed boundary (IB) method to simulate platelet motion and deformation in shear flow. The kinetic-based adhesive dynamics model was also integrated into the three dimensional model to simulate formation and disassociation of the receptor-ligand bonds during the platelet-platelet and platelet-vessel wall interactions. Parallelized model simulations were implemented on a GPU computer cluster which speeded up simulations by a factor of 100 (see **Table 3**) in comparison with CPU implementation which allowed for the first time to run biologically relevant predictive simulations.

By using novel biologically calibrated 3D modeling approach, it is shown that the platelet stiffness, the number of GPIbα platelet functional receptors and mutual interaction between platelets can significantly alter the adherence of platelets at the site of vascular injury. Our results demonstrate how a comprehensive modeling approach coupling three biologically relevant scales can provide new insights into the biomedically important problem of early thrombus development. Variation of the number of functional GPIbα platelet receptors as well as changes of platelet stiffness can represent effects of specific drugs for reducing or enhancing platelet activity. This emphasizes the importance of predictive simulations as it can potentially improve the search for new drug targets and help with making treatment of thrombosis patient specific.

Damage or alteration of a blood vessel lining can result in activation of flowing platelets and their subsequent aggregation at sites of vascular injury. The ability of platelets to tether to and translocate on injured vascular endothelium relies on the interaction between the platelet glycoprotein receptor

Ibα(GPIbα) and the A1 domain of von Willebrand factor (vWF-A1) [1].

Along with biochemical activation of platelets, large shear disturbances of blood flow is one of the key factors promoting pathogenic activation of platelets and formation of thrombi. In addition, platelets flowing in a whole blood exhibit increased concentrations in the vicinity of the vessel wall, making platelet-platelet interactions more frequent near vascular surfaces. Excessive accumulation of platelets at injury sites is one of the pathological events that result in acute myocardial infarction, sudden death, and ischemic stroke. This pathological process is responsible for mortality and morbidity rates higher than for any other disease, making platelet a major target for therapeutic interventions. Thus, studying an individual platelet dynamics as well as platelet-platelet interactions and platelet adhesion to a vascular or thrombus surface is of high biomedical importance and urgency.

Effects of shear flow on accumulation of platelets on various surfaces have been extensively studied in *in vitro* and *in vivo* experiments [1-5]. However, there is a limited amount of available experimental data on an individual platelet dynamics in the vicinity of the vascular surface as well as platelet-surface attachment. There is also a lack of experimental data demonstrating how platelet-surface attachment is affected by mechanical properties of a platelet as well as by platelet receptor-ligand kinetics. Better understanding of platelet aggregation requires study of the interplay among biochemical, mechanical and hydrodynamic processes occurring at different scales, including a nm-scale (receptor-ligand kinetics), a μm-scale (cellular level), and a mm-scale (early platelet aggregate). Multiple characteristic scales make it difficult to experimentally discern effects of different processes involved in platelet-surface attachment and overall thrombus growth dynamics. Meanwhile, a multiscale modeling approach can provide a useful predictive tool to aid in elucidating mechanisms of platelet-wall attachment and aggregation.

Several multiscale models attempting to couple large number of submodels at different scales have been developed (see, amongst others, for reviews [6-7]). These models implemented simplified submodels in order to make simulations less computationally expensive. It is extremely difficult at this time, if not impossible, to validate predictions of multiscale models attempting to combine submodels at all scales representing processes of blood clot formation using existing experimental data. Also, most experimental data are currently available at the molecular level and individual platelet level. Therefore, it is important to develop detailed multiscale models coupling two or three scales and considering only a few processes at a time. Such models when properly calibrated with available experimental data can provide useful predictive tools aiding in designing new experiments, drug design and planning new patient specific therapeutic strategies.

Several computational models have been developed to characterize platelet and other types of blood cells motion and adhesion dynamic under hydrodynamic shear flow at cell- and receptor-levels (see [6-7] for a review). Analytical solutions for forces and torques exerted on a platelet treated as a rigid object in Stokes regime in a 2D case were obtained in [8] and compared with the data obtained using image analysis algorithm for tracking the motion of platelets before, during, and after contact with the surface. Kinetic properties of the receptor-ligand adhesion bonds, GPIbα-von Willebrand Factor (vWF), were quantified in [1] and [4] using Monte Carlo simulations and pause time analysis of transient capture/release events. This approach provided association and disassociation rate constants $k_{on}$ and $k_{off}$, depending on the shear rate of the blood flow.

Experimental study in [9] showed that platelets have viscoelastic properties and the elastic moduli in the range of 1 to 50 KPa. Large deformation occurred when platelets were suspended in the shear flow [10]. To account for the elastic and viscoelastic properties of cells, a number of methods accounting for cell structural properties have been developed [11-13]. The subcellular element (SCE) model introduced in Sandersius & Newman [11] represented each cell by a collection of elastically coupled SCEs, interacting with each other via short-range potentials. Sweet *et al* [12] and Xu *et al* [13] presented a 3D modeling approach in which cells, modeled by SCEs, were coupled with fluid flow and substrate models by using Langevin equation.

The fluid-structure interaction approach is an essential part of the model. Previously, the immersed boundary method (IBM) was introduced by Peskin [35] to investigate the blood flow in the human heart, has been applied to many other fluid–structure interaction problems, including platelet aggregation [47] and deformation of red blood cells [48]. Skorczewski *et al* [49] developed a two-dimensional model using a lattice-Boltzmann immersed boundary method to investigate the motion of platelets near a vessel wall and close to an intravascular thrombus, in which they modeled the platelets as rigid bodies while the red blood cells were represented as deformable bodies.

The results of the predictive simulations of the 3D model introduced in this paper revealed that the platelet pause time strongly depends on the stiffness of the platelet as well as on the number of expressed GPIb membrane functional receptors. Additionally, we demonstrated that the platelet-platelet interaction near the surface of the vessel wall could significantly decrease the platelet paused time, and thus decrease the rate of platelet attachment to the injury site.

The paper is organized as follows. It starts with the Biological background section. Then, methodological innovation is described in detail including description of submodel at each of three space scales and of the coupling approach. This is followed by the Results section which includes model validation and description of the predictive simulation. Biological relevance of the predictive simulations is discussed in the Discussion section. GPU implementation of the 3D model is described in the Appendix.

## 2. Biological background

The mechanism by which platelets bind to a damaged blood vessel wall is similar to that of leukocyte binding to activated endothelium [18], and requires two binding steps. The first step is the rapid formation of unstable catch-slip bonds which slow platelet and cause platelet flipping along the damaged surface. (Counter intuitively, the dissociation rate first decreases with increasing force until reaching a threshold.) This is mediated by the platelet receptor component, GPIbα, forming transient bonds with the von Willebrand Factor (vWF) exposed at the injury site. Rapid association and dissociation kinetics of the bonds result in transient tethering and subsequent flipping (or rolling) and pausing of platelets on the vessel surface [1, 19]. Then, stable bonds slowly form between platelet receptors and ligands (often integrin αIIbβ3 binding with vWF or fibrinogen) bound to the damaged wall or the surface of the thrombus resulting in strong adhesion, initiating transmembrane and, subsequently, intracellular signaling. As the blood clot grows, platelet-platelet interaction becomes one of the major factor determining clot growth rate and integrity as platelets expose GPIIbIIIa receptors which permit platelet-platelet adhesion via fibrinogen. Adhesion of platelets to the injured surface is also affected by shear rates of the flow. At high shear, platelet integrin α2β1 and GPVI receptors are not sufficient to initiate binding to collagen, and binding of the GPIbα receptor to vWF immobilized on collagen, becomes essential in platelet adhesion.

The stiffness of the platelet not only determines the shape and morphology of the clot but also affects clot mechanical properties; as platelet stiffness determines cell shape when exposed to various flow conditions and contact interaction with other cells and blood vessel wall. This will affect the number of receptor-ligand pairs in platelet-platelet and platelet-substrate interactions. Platelet stiffness is also an important property reflecting platelet functioning, since it reorganizes its structure during activation or as a response to physiological or pathological conditions.

To date, to the best of our knowledge, cumulative effects of platelet stiffness, different levels of expression of GPIbα receptors and platelet-platelet interaction impacting strength of platelet-substrate binding have not been systematically investigated. Our model provides a unique means for quantitatively understanding these effects, which are critical for improving our knowledge about the initial stage of the blood clot formation.

## 3. Methodological innovation of the three-dimensional modeling approach

The novelty of the three-dimensional (3D) model lies in developing novel membrane submodel as well as in new approaches of coupling submodels of biological processes at three spatial scales (see **Figure 1**) which are crucial to early blood clot formation. At the subcellular scale (nano-scale), a kinetic-based stochastic dynamic adhesion submodel is used to simulate vWF-GPIbα binding and GPIbα-vWF-GPIbα binding, in which individual vWF and GPIbα molecules are represented as elastic springs. This is justified by the fact that these receptor-ligand binding is probabilistic in nature [1]. Moreover, individual filaments in the cytoskeleton network of the platelet membrane are treated as coarse-grained harmonic springs. At the cellular scale, a novel continuum description of the lipid bilayer of the cell membrane is utilized. We developed this new platelet membrane model to study effects of membrane stiffness on cell-substrate interaction, which was shown to strongly affect platelet-injury site adhesion. (See also subsection *4.3.1* for model prediction.) The subcellular scale and the cellular scale components are integrated by distributing GPIbα receptors at the vertices of the cytoskeleton network and by superimposing the cytoskeleton network and the lipid bilayer. At the macro scale, the dynamics of the fluid flow is represented using the LBM to facilitate parallelizing the simulation code on GPUs. The platelet model is coupled with the LBM using the IBM. (The coupling and data flow between all the submodels are demonstrated in **Figure** 1.) We calibrate and validate this 3D model by comparing simulations at different scales with either theoretical results or available experimental data at these scales. Specifically, the platelet model coupling with LBM was validated using theoretical results and previous simulation results (See also Section 4.1); while the platelet-substrate adhesion simulations were compared with experimental data to calibrate the DAM submodel under different flow conditions (See also Section 4.2).

At each time step of simulation, the hybrid membrane model is first used to calculate forces acting on the nodes of the Lagrangian mesh representing platelet geometry, such as bond forces resulting from stretching or compression of cytoskeleton network, bending forces resulting from deformation of the lipid bilayer and attraction/repulsion between platelet and environment due to formed ligand-receptor bonds. This is followed by coupled LBM and IBM to update fluid flow and position of platelet. Finally, the MC computations of platelet adhesion to a surface expressing vWFs are performed to break the already formed bonds and to generate new bonds from unbound GPIBα and vWF.

We note that this is the first time that a detailed platelet membrane model has been developed and implemented on GPUs for studying cell-flow, cell-cell and cell-substrate interactions. Because of the speedup gained by GPU implementation, we are able to investigate effects of these interactions and cell mechanics on platelet dynamics in a timely manner. Additionally, this model can be directly used for modeling any biological cells with membrane structures similar to those of eukaryotic cells. In this section we describe in detail individual submodels and explain how they are coupled.

## 3.1 Platelet membrane submodel

We simulate the motion of platelets in a 3D region bounded by an infinite flat plane at z = 0 (see **Figure 2a** for example). A platelet has initial shape defined by $\frac{x^2}{a^2} + \frac{y^2}{a^2} + \frac{z^2}{(\lambda a)^2} = 1$, where a = 1 μm is the approximate particle radius and $\lambda = 0.25$ aspect ratio [20-21]. The Reynolds number of this system is Re = $\gamma\rho a^2/\mu = O(10^{-3})$, where $\gamma = 300$ and $400\ s^{-1}$ are the shear rates used in experiments [1], a = 1 μm is the particle radius, $\rho = 1.0239\ g/cm^3$ is the density of blood plasma, and μ = 1.2 cP is the viscosity of blood plasma [22].

The platelet membrane, which is similar to the membrane of a red blood cell, is also assumed to consist of a lipid bilayer and an attached cytoskeleton. Following ideas from [23] and [16], the platelet membrane surface geometry is represented by a triangular mesh consisting of a collection of N (N = 958 in our simulation) points $\{\boldsymbol{X}_i, i \in 1 \dots N\}$ (see **Figure 2b**). The connected edges of the mesh are used to model the cytoskeleton network of the platelet membrane and the triangulated mesh surface represents the lipid bilayer of the cell membrane, where the cytoskeleton attaches to. The mesh points represent coarse-grained actin vertices and each edge of the mesh represents a coarse-grained filament. The Helmholtz free energy of the membrane is defined to be

$$H_{membrane} = H_{SCE} + H_{bending} + H_{volume} + H_{area} + H_{wall} \qquad (1)$$

Here, term $H_{SCE}$ is the in-plane energy of the cytoskeleton network; $H_{bending}$ is the bending energy representing the bending resistance of the lipid bilayer; $H_{volume}$ and $H_{area}$ are volume, area conservation constraints, respectively; and $H_{wall}$ represents the energy relating to interactions due to ligand-receptor binding (explained in detail in Section 3.3).

We employ a harmonic 'spring' model to simulate the elasticity of the edge connecting mesh points $i$ and $j$, which mimics a coarse-grained filament. The associated potential energy functions for points $i$ and $j$ are

$$U_{ij}^e = \frac{k}{2}\left(\left\|\boldsymbol{R}_{ij}\right\| - L_{ij}\right)^2 \qquad (2)$$

where $L_{ij}$ is the rest length, $\boldsymbol{R}_{ij} = \boldsymbol{X}_j - \boldsymbol{X}_i$ the position vector difference for points $i$ and $j$, respectively, and $k = 2E\Delta x/5$ the coefficient that defines the spring 'stiffness' [24] for elastic modulus $E = 25\ kPa$ [9] and $\Delta x = 0.1\ \mu m$ unit link length of the spring. The total potential energy for the cytoskeleton network is $H_{SCE} = \sum_{i,j\ for\ all\ edges} U_{ij}^e$. The corresponding force vector acting on point $i$ by point $j$ is

$$F_{ij}^e = -\nabla_x U_{ij}^e(x) = -k\left(\left\|\boldsymbol{R}_{ij}\right\| - L_{ij}\right)\frac{\boldsymbol{R}_{ij}}{\left\|\boldsymbol{R}_{ij}\right\|} \qquad (3)$$

The area and volume conservation constraints, which account for area incompressibility of the lipid bilayer and incompressibility of the inner cytosol, respectively, are expressed as

$$H_{area} = \frac{k_s\left(S^{total} - S_0^{total}\right)^2}{2S_0^{total}} + \sum_{all\ triangles}\frac{k_t(S - S_0)^2}{2S_0} \qquad (4)$$

$$H_{volume} = \frac{k_v(V - V_0)^2}{2V_0} \qquad (5)$$

where $k_s$, $k_t$, and $k_v$ are the global area, local area, and volume constraint coefficients, respectively. The terms $S^{total}$ and $V$ denote the surface area and volume for the whole platelet, while $S_0$, $S_0^{total}$ and $V_0$ are the individual triangle mesh area, the total membrane area and the volume for unstressed platelet, respectively.

We adopt the energetic variational approach developed in [15] to represent the lipid bilayer of the cell membrane. Let $\Sigma \in R^3$ be a smooth, closed surface representing the lipid bilayer of the platelet. The bending energy of the lipid bilayer is defined as [15]:

$$H_{bending} = k_0 \int_\Sigma \frac{1}{2} K(\mathbf{x})^2 dS(\mathbf{x}) \qquad (6)$$

where $K(x) = \frac{1}{2}\big(\kappa_1(\mathbf{x}) + \kappa_2(\mathbf{x})\big)$ is the mean curvature, and $\kappa_1(\mathbf{x})$, $\kappa_2(\mathbf{x})$ are the principle curvatures at the point $x$. We follow the finite element method in [25] to calculate $\kappa_1(\mathbf{x})$ and $\kappa_2(\mathbf{x})$. Briefly, let $u(\xi, \eta)$ be a function defined over a triangle of the surface mesh representing the lipid bilayer and approximated as

$$u(\xi, \eta) = \sum_{i=1}^{6} u_i N_i(\xi, \eta) \qquad (7)$$

where $\xi$, $\eta$ are the local parametric coordinates, $u_i$ is the value of u at node $i$ and $N_i(\xi, \eta)$ are the basis functions for a quadratic six-node triangular finite element. To evaluate the membrane curvature tensor $\boldsymbol{\kappa}$, one needs to calculate the left Cauchy–Green strain tensor, which is determined from the surface deformation gradient tensor, $\boldsymbol{A}$. For each triangular element, the surface deformation gradient tensors at the element nodes are obtained by solving the following system of equations,

$$\boldsymbol{A} \cdot \frac{\partial \bar{X}}{\partial \xi} = \frac{\partial \boldsymbol{X}}{\partial \xi}, \ \boldsymbol{A} \cdot \frac{\partial \bar{X}}{\partial \eta} = \frac{\partial \boldsymbol{X}}{\partial \eta}, \ \boldsymbol{A} \cdot \bar{\boldsymbol{n}} = \boldsymbol{0} \qquad (8)$$

at each node of the element, and $\boldsymbol{X}$, $\bar{\boldsymbol{X}}$ are its positions in the unstressed state and after deformation at time t, respectively. $\bar{\boldsymbol{n}}$ is the unit normal vector to the un-deformed membranes. To evaluate the curvature tensor $\boldsymbol{\kappa}$ at a point, one needs to solve

$$\frac{\partial \boldsymbol{X}}{\partial \xi} \cdot \boldsymbol{\kappa} = \frac{\partial \boldsymbol{n}}{\partial \xi}, \ \frac{\partial \boldsymbol{X}}{\partial \eta} \cdot \boldsymbol{\kappa} = \frac{\partial \boldsymbol{n}}{\partial \eta}, \ \boldsymbol{n} \cdot \boldsymbol{\kappa} = \boldsymbol{0} \qquad (9)$$

at each element node and then average over the elements sharing that node, and $\boldsymbol{n}$ is the unit normal vector to the deformed membranes. The mean curvature is given by

$$K(x) = \frac{1}{2}(\kappa_1 + \kappa_2) = \frac{1}{2} tr(\boldsymbol{\kappa}) \qquad (10)$$

The normal component of the elastic force associated bending energy (6) is obtained by taking variational derivative and is given by $\boldsymbol{F}_{bend} = (\Delta_\Sigma K + 2K^3)\boldsymbol{n}.$

Thus, nodal forces $\boldsymbol{F_i}$ are derived from the total energy as follows

$$\boldsymbol{F}_i = \frac{\partial(H_{SCE} + H_{volume} + H_{area} + H_{wall})}{\partial x_i} + \boldsymbol{F}_{bend} \qquad (11)$$

Computation of $\frac{\partial(H_{wall})}{\partial x_i}$ is explained in **Sec. 3.3**.

## 3.2 Platelet stochastic dynamic adhesion submodel (DAM)

The kinetic-based stochastic dynamic adhesion submodel based on ideas of the Dembo Model [16-17] is used to simulate the GPIbα−un-activated platelet binding to immobilized vWF on the vessel wall or platelet-platelet adhesion through forming GPIbα−vWF−GPIbα bonds, in which vWF was originally in plasma. Here, we provide details of the model for GPIbα−(immobilized) vWF bond formation; modeling of GPIbα−vWF−GPIbα is treated similarly. Each platelet has approximately 10,688 GPIbα receptors distributed uniformly on its membrane surface, to achieve a surface density of ~1500 receptors/$\mu m^2$ [26]. In our model, 5344 receptor point locations on the platelet surface are randomly distributed on the platelet membrane mesh, with each point location representing two GPIbα receptors, since there are two GPIbα receptors existing on each GPV molecule [27]. On the bottom plane of the simulation domain, z = 0, immobilized vWFs are uniformly distributed, resulting in a vWF density of 25 $\mu m^{-2}$, which is consistent with experimental conditions in Doggett *et al* [1].

The following rules are used for governing the GPIbα−vWF binding [28]. 1) Two vWF molecules cannot bind to the same receptor nodes for reasons of steric blocking; and 2) receptors from a maximum of 4 receptor nodes present on a platelet surface can bind a vWF molecule.

In our stochastic DAM, when an unbound GPIbα and an unbound vWF are separated less than the length of GPIbα−vWF bond of 128 nm [29-30], test for forming a bond is performed. Next, the formed bonds are tested for breakage. A GPIbα-vWF bond is modeled as a linear spring.

Probabilities of GPIbα−vWF bond formation and dissociation are calculated using $P_f$ (probability of forward reaction) and $P_r$ (probability of reverse reaction) described in [31]: $P_f = 1 - \exp(-k_{on}\Delta t)$, $P_r = 1 - \exp(-k_{off}\Delta t)$, where $k_{off}$ and $k_{on}$ are given in s$^{-1}$ units and $\Delta t$ is the simulation time step. The reverse rate constant is calculated using the Bell model for force dependent dissociation rate of weak noncovalent bonds:

$$k_{off} = k_{off}^0 \exp\left(\frac{\gamma F_b}{k_B T}\right) \qquad (12)$$

where $k_{off}(F_b)$ is the bond dissociation rate, $k_{off}^0$ is the unstressed off-rate, $\gamma$ is the reactive compliance, $F_b$ is the applied force on the bond, and $k_B T$ is the product of Boltzmann constant and temperature. The dependence of bond formation rate constant $k_{on}$ on the deviation bond length is described by [16, 28] as

$$k_{on} = k_{on}^0 \exp\left(\sigma|x_b - l_b|\frac{\gamma - 0.5|x_b - l_b|}{k_B T}\right) \qquad (13)$$

where $k_{on}^0$ is the intrinsic cross-linking formation rate constant, $\sigma$ is the spring constant, $l_b$ is the equilibrium bond length, $x_b$ is the distance spanning the endpoints of the GPIbα receptor on the platelet surface and the vWF-A1 binding site.

The adhesion force of the GPIbα-vWF bond located at *i*th node of the cell membrane is calculated using a spring model as follows,

$$F_{bond,i} = -\sigma|x_b - l_b| \qquad (14)$$

where $\sigma$ is the spring constant, $l_b$ is the equilibrium bond length, $x_b$ is the distance spanning the endpoints of the GPIbα receptor on the platelet surface and the vWF-A1 binding site. **Table 2** lists values of parameters of the DAM used in simulations.

### 3.3 Lattice Boltzmann method for simulating blood flow

The LBM employs purely localized fluid particle evolution and relaxation, which in turn facilitates parallelization in computer implementation. The LBM decomposes the fluid domain into structured lattice nodes and operates on the lattice. The fluid is modeled as a group of fluid particles that are only allowed to move between lattice nodes or stay at rest. The composition of the lattice nodes depends on the chosen lattice model. In this paper, we used the 3D model of a cubic lattice ($16 \times 64 \times 16$ μm with spacing $h = 0.2$ μm) with 19 discrete velocity directions (model D3Q19, as shown in **Figure** 4). The LBM solves the Boltzmann equation describing the dynamics of fluid from a microscopic point of view: in fluid, particles with velocities $\boldsymbol{v}_i$, collide with certain probability and exchange momentum. The collisions are assumed to be ideal, that is the total momentum and energy is conserved during the collisions. The Boltzmann equation describes probability $\boldsymbol{f}(\boldsymbol{x}, \boldsymbol{v}, t)$ of finding a particle with velocity $\boldsymbol{v}$ at a position $\boldsymbol{x}$ and at time $t$ evolves with time:

$$\boldsymbol{v} \cdot \nabla_x f + \boldsymbol{F} \cdot \nabla_p f + \frac{\partial f}{\partial t} = \Omega(f) \qquad (15)$$

where $\boldsymbol{F}$ denotes an external body force, $\nabla_{x,p}$ is the gradient in position and momentum space, and $\Omega(f)$ denotes collision operator which is chosen as a relaxation of $f$ with a characteristic time $\tau$ to the equilibrium distribution $f^{(eq)}(\boldsymbol{v}, \rho)$:

$$\Omega(f) = -\frac{1}{\tau}(f - f^{(eq)}) \qquad (16)$$

The equilibrium distribution function depends on the local density $\rho(\boldsymbol{x}, t)$ and the velocity field $\boldsymbol{v}(\boldsymbol{x}, t)$. In D3Q19 lattice model, 19 values $f_i(\boldsymbol{x}, t)$ are stored at each lattice site assigned to a lattice vector $\boldsymbol{c}_i$. The local density at a lattice point are obtained by summing all $f_i$,

$$\rho(\boldsymbol{x}, t) = \sum_{i=1}^{19} f_i(\boldsymbol{x}, t) \qquad (17)$$

and the streaming velocity is given by

$$u(\boldsymbol{x}, t) = \frac{1}{\rho(\boldsymbol{x}, t)} \sum_{i=1}^{19} f_i(\boldsymbol{x}, t)\boldsymbol{c}_i \qquad (18)$$

where $\boldsymbol{c}_i = h/\Delta t$ is the lattice speed associated with the *i*th direction and $\Delta t$ is the time step of our simulation.

Using a Chapman Enskog expansion, Guo et al. [32] showed that the following lattice Boltzmann equations give a second-order-accurate $\boldsymbol{v}$, the Navier–Stokes velocity in the presence of a spatially varying, time-dependent force:

$$f_i(\boldsymbol{x} + \boldsymbol{c}_i \Delta t, t + \Delta t) = f_i(\boldsymbol{x}, t) - \frac{1}{\tau}\Big(f_i(\boldsymbol{x}, t) - f_i^{eq}(\rho, \boldsymbol{v})\Big)$$

$$+ \omega_i \Delta t \left(1 - \frac{1}{2\tau}\right)\left[\frac{(\boldsymbol{F} \cdot \boldsymbol{c}_i)}{c_s^2} + \frac{(\boldsymbol{u}\boldsymbol{F}^T + \boldsymbol{F}\boldsymbol{u}^T):(c_i c_i^T + c_s^2 I)}{2c_s^4}\right] \qquad (19)$$

where $\boldsymbol{u}$ is a streaming velocity defined in Eq. (18), $\boldsymbol{v} = \boldsymbol{u} + \boldsymbol{F}\Delta t/2\rho$, and

$$f_i^{(eq)}(\rho, \boldsymbol{v}) = \omega_i \rho \left[1 + \frac{\boldsymbol{c}_i \cdot \boldsymbol{v}}{c_s^2} + \frac{(\boldsymbol{c}_i \cdot \boldsymbol{v})^2}{2c_s^4} - \frac{v^2}{2c_s^2}\right] \qquad (20)$$

with the lattice speed of sound $c_s = \frac{1}{\sqrt{3}}h/\Delta t$ for the D3Q19 lattice and the lattice weights

$$\omega_i = \begin{cases} 2/36 & i = 1\dots6, \\ 1/36 & i = 7\dots18, \\ 12/36 & i = 19. \end{cases} \qquad (21)$$

The pressure $p = c_s^2 \rho$ turns out to be proportional to the density and the dynamic shear viscosity is given by

$$\eta = c_s^2 \rho \left(\tau - \frac{1}{2}\right) \qquad (22)$$

To ensure convergence and stability of LBM, we follow the method in [33] to choose our parameters. Spacing $h = 0.2\,\mu m$ was determined by our simulated fluid domain and memory size of the GPU card. Time step $\Delta t$ was determined from the equation [33]: $\Delta t = (\tau - 0.5)h^2/(3\upsilon)$, where $\upsilon = \mu/\rho$ is kinetic viscosity, $\mu$ and $\rho$ are fluid viscosity and density as defined in **Table 2**. Generally speaking, a larger value of $\tau$ leads to a more stable LBM simulation, and $\tau$ must be greater than 0.5. We set $\tau = 1.379\,s$ in our model, such that $\Delta t = 10^{-8}s$.

Periodic boundary conditions in x-z and y-z boundary planes (y = 0, y = 64 μm, x = 0 and x = 16 μm), are realized by propagating the $f_i$ from the computational domain on the one boundary to the boundary on the opposite side of the domain. In the x-y boundary planes we used the on-site velocity boundary conditions proposed by Hecht and Harting [34]. For instance, in x-y boundary plane z = 0, $f_i$ ($i = 1, 2, 3, 4, 6, 7, 8, 10, 11, 12, 14, 16, 18, 19$) can be obtained from the streaming step, but $f_i$ ($i = 5, 9, 13, 15, 17$) are undetermined. Following the methods of Hecht and Harting [34], we can get

$$\rho = \frac{1}{1 - v_z}[f_1 + f_2 + f_3 + f_4 + f_7 + f_8 + f_{11} + f_{12} + f_{19} + 2(f_6 + f_{10} + f_{14} + f_{16} + f_{18})] \qquad (23)$$

$$f_5 = f_6 + \frac{1}{3}\rho v_z \qquad (24)$$

$$f_9 = f_{14} + \frac{\rho}{6}(v_z + v_x) - N_x^z \qquad (25)$$

$$f_{13} = f_{10} + \frac{\rho}{6}(v_z - v_x) + N_x^z \qquad (26)$$

$$f_{15} = f_{18} + \frac{\rho}{6}(v_z + v_y) - N_y^z \qquad (27)$$

$$f_{17} = f_{16} + \frac{\rho}{6}(v_z - v_y) + N_y^z \qquad (28)$$

Here, $v_x$, $v_y$ and $v_z$ are boundary velocities in x, y and z directions, $N_x^z$ and $N_y^z$ the transverse momentum corrections on the z-boundary for distributions propagating in x-and y directions, respectively:

$$N_x^z = \frac{1}{2}[f_1 + f_7 + f_8 - (f_2 + f_{11} + f_{12})] - \frac{1}{3}\rho v_x \qquad (29)$$

$$N_y^z = \frac{1}{2}[f_3 + f_7 + f_{11} - (f_4 + f_8 + f_{12})] - \frac{1}{3}\rho v_y \qquad (30)$$

### 3.4 Coupling platalet, DAM and flow submodels

3.4.1 Coupling platelet and DAM submodels

As described in **Sec. 3.3**, GPIbα receptors are randomly distributed on the cell membrane. In each step of the simulation, forming and breaking a GPIbα-vWF bond is updated using the DAM. When a formed bond is either stretched or compressed, the bond deformation force is computed using Eq. (14). This force is exerted on the cell membrane at the place where the GPIbα receptor of the bond is located. When only the platelet membrane and vessel wall interaction is considered, the term $H_{wall}$ of Eq. (1) represents the energy associated with these interactions. In particular, $\frac{\partial(H_{wall})}{\partial x_i}$ corresponds to the sum of following two forces: 1) adhesion forces caused by GPIbα-vWF bond and 2) short range repulsive forces accounting for contact of vessel wall. The short range repulsive force is given by an empirical relationship as: $F_{rep} = F_0 \frac{\tau e^{-\tau\varepsilon}}{1 - e^{-\tau\varepsilon}}$ , where $F_0 = 500\ pN \cdot m$, $\tau = 2000\ \mu m^{-1}$ and $\varepsilon$ is the separation distance between platelet membrane and vessel wall [16]. Thus, $\frac{\partial H_{wall}}{\partial x_i}$ term in Eq. (11) is defined to be:

$$\frac{\partial H_{wall}}{\partial x_i} = F_{bond,i} + F_{rep}$$

3.4.2 Coupling cell and flow submodels

To couple the integrated platelet and stochastic DAM submodels with the blood flow computed by LBM, we utilize the IBM [35]. In the IBM (**Figure** 3) Eulerian description is used for the fluid dynamics, and Lagrangian description is used for objects immersed in the fluid. Using lowercase letters for Eulerian variables, and uppercase letters for Lagrangian variables, we have

$$\frac{dX}{dt} = U(X,t) = \int_{\Omega_f} u(x,t)\delta(x-X)dx \qquad (31)$$

$$f(X,t) = \int_{\Gamma_b} F(X,t)\delta(x-X)dX \qquad (32)$$

where $t$ is time, $\boldsymbol{u}$ the flow velocity, $\boldsymbol{U}$ the speed of the solid object boundary, $\boldsymbol{x}$ the fluid flow coordinate, $\boldsymbol{X}$ the boundary coordinate, $\boldsymbol{f}$ the force density on the fluid node, $\boldsymbol{F}$ the force density on the solid elements and $\delta(\boldsymbol{r})$ the Dirac delta function.

Eqs. (31) and (32) are approximated using a regularized discrete delta function $\delta_h$. The discretized forms of Eqs. (31) and (32) using $\delta_h$ are as follows

$$\frac{dX_m}{dt} = U_k = \sum_{i,j,k} \boldsymbol{u}_{ijk} \; \delta_h\big(\boldsymbol{x}_{ijk} - \boldsymbol{X}_m\big)h^3 \qquad (33)$$

$$\boldsymbol{f}_{ijk} = \sum_m \boldsymbol{F}_m \, \delta_h\big(\boldsymbol{x}_{ijk} - \boldsymbol{X}_m\big)h^3 \qquad (34)$$

where $h$ is the fluid node spacing, $\boldsymbol{x}_{ijk} = (ih, jh, kh)$ the coordinate of the $i, j, k$th Eulerian grid node, $\boldsymbol{X}_m$ the Lagrange coordinate of the $m$th elements, $\boldsymbol{f}_{ijk}$ the force density on $\boldsymbol{x}_{ijk}$, $\boldsymbol{F}_m$ the force density on $\boldsymbol{X}_m$, $\boldsymbol{u}_{ijk}$ the velocity of $\boldsymbol{x}_{ijk}$, $\boldsymbol{U}_k$ the velocity of $\boldsymbol{X}_m$. The discrete delta function $\delta_h$ appearing in Eqs. (33) and (34) is a smoothed approximation to the Dirac delta function $\delta(\boldsymbol{r})$. (The detailed derivation procedures in several forms were presented in literature [36].) We use the following common form

$$\delta_h(x, y, z) = \frac{1}{h^3} \phi\left(\frac{x}{h}\right) \phi\left(\frac{y}{h}\right) \phi\left(\frac{z}{h}\right) \qquad (35)$$

$$\phi(x) = \begin{cases} \frac{1}{4}\left(1 + cos\left(\frac{\pi x}{2}\right)\right) & for \; 0 \leq |x| \leq 2 \\ 0 & for \; |x| \leq 2 \end{cases} \qquad (36)$$

To sum up, firstly, in each step of the simulation, Eq. (20) is solved. Then, positions of nodes of the platelet membrane are updated by Eq. (33). Finally, the MC computations are performed to break the already formed bonds and to generate new bonds from unbound GPIbα and vWF.

## 4. Results

First, the model was verified by comparing simulation results with analytical solutions and available model simulation data [22]. Next, we validated the model by comparing the simulation results with the experimental data [1] on flipping platelets flowing over a vWF-coated surface. Calibrated 3D model was used to predict how the stiffness of a platelet membrane, the number of receptors on platelet membrane and the strength of platelet-platelet adhesion affect the paused time of the platelet adhering to a vessel wall.

### 4.1 Validation of fluid-platelet coupling by comparing with the Jeffery orbit

Mody *et al.* [8] described theoretical solutions using the Jeffery orbit theory and provided predictions obtained using the analytical platelet-flipping model. This analytical solution (shown as solid red line in **Figure 6**) did not consider the wall effect and only applied to the cases of platelet motion far from the wall ($H/a > 20$) [22], where $H$ is the centroid height of platelet and a is the major radius (as shown in Figure 2). Mody *et al.* [22] modified the completed double layer-boundary integral equation method to include a flat surface boundary that was used to compute the effects of the wall on the flow behavior of a platelet. Platelets located as far as 2.4-fold platelet radii from the surface display "modified" Jeffery orbits with periodic rotational motion in the direction of flow (green dash line in **Figure 6**). To verify our model, we simulated the flipping of a single platelet located at the distance of 2.4*a* as well as greater than 20*a*, from the vessel wall. Our simulations revealed that the calculated orbit of rotation (blue dash line in **Figure 6**) agreed well with the Mody's simulation results [22] (green dash line) within an experimental error of 2.65% for platelet located at the distance of 2.4*a,* and agreed perfectly with Mody's simulation results [22] for platelet located at the distance > 20*a*. **Figure 5** shows the series of snapshots from our simulations of a platelet flipping in a shear flow near the vessel wall. . In our model the platelet was modeled as an elastic cell with the elastic modulus measured by the AFM experiments [9]; while Mody *et al.* [8] considered the platelet as a rigid object. By comparing our simulation results and results of Mody *et al.* [8], we conclude that our simulations can be successfully implemented to model the motion of individual resting platelets revealing high stiffness membrane values.

### 4.2 Validation of the model of the platelet-substrate adhesion

To validate the kinetic submodel, we simulated flowing platelets adhering to substrate through GPIbα-vWF binding and calculated $k_{off}$ rates to compare with available experimental data. The model parameters used in our simulations (**Table 2**) were obtained in biological experiments [1, 9, 15, 20-23]. The adhesive dynamic parameters were measured in *in vitro* flow chamber tests [1].

Doggett et al. [1] measured the kinetics that governs platelet interactions with vWF in hemodynamic flow. In their experiment, the frequency of tethering for platelets was measured by determining the percentage of cells that paused, but did not translocate, on vWF substrates. The frequency of tethering for microspheres coated with vWF on antibody-immobilized platelet substrates was also measured. A transient tether event was defined as flowing platelet that abruptly halted forward motion for a defined period of time and subsequently released, without evidence of translocation, to resume a velocity equivalent to that of a non-interacting cell. Dissociation rate constants ($k_{off}$) were determined by plotting the natural log of the number of beads that interacted as a function of pause time after the initiation of tethering (**Figure 7**, the slope of the line is $-k_{off}$).

To calculate dissociation constants, we performed simulations for various numbers of random seeds (1000-1200). The results of the simulations and experimental data are presented in **Figure 7** for two different flow shear rates as the natural log of the number of platelets tethering events versus the pause time. The values of dissociation rate $k_{off}$ were found to be 3.31 and 3.58 s[-1] for flow shear stress 3.0 and 4.0 dyn/cm[2] respectively. The corresponding experimental values obtained in [1] were 4.83 and 5.18 s[-1].

It should be mentioned that our simulations confirmed several experimental observations. It was reported in [1] that in the range of flow rates considered, forces acting on the GPIbα-vWF bond were not sufficient to alter the rate of dissociation $k_{off}$. Our simulations also demonstrated that the $k_{off}$ values only altered in a small range from 3.31 to 3.58 s$^{-1}$. Additionally, it was reported in [1] that the forces acting on a platelet in shear flow were 14.7 and 19.6 pN for flow shear stress 3.0 and 4.0 dyn/cm$^2$ respectively, while our model yielded very close force values of 12.8 and 15.6 pN, respectively.

Our simulations also confirmed that in the range of flow rates (0 – 4 dyn/cm$^2$ wall shear stress), bond association and dissociation kinetics can be successfully described by Dembo Model (Eq. (12, 13)).

**4.3 Predictive simulations**

The responses of a platelet to interactions with environment depend, among others, on the mechanical forces that platelets experience. In this section, we consider effects of platelet membrane tension, flow shear stresses, and adhesion bond forces on platelet-substrate adhesion dynamics.

*4.3.1 Effect of platelet membrane stiffness*

Simulation results in Section 4.1 show that the flow dynamics of the platelet in linear shear flow can be studied by modeling platelet as rigid objects. How the stiffness of the platelets affects the platelet-substrate interaction remain to be answered. In [39], it was reported that alteration of platelet stiffness can modulate platelet aggregation. We hypothesized that softer cells lead to prolonged adhesion time and could potentially increase chances of platelets to be activated after adhesion. Here, we report the simulation results indicating remarkable changes in platelet paused time as the platelet membrane stiffness changes. We varied the platelet membrane stiffness from 25 KPa to 2.5 KPa, and performed simulations with 30 different random seeds to obtain 30 different paused times under flow shear stress of 3.0 dyn/cm$^2$. The paused time was 6.69 ± 0.71s (M ± SD) for the membrane stiffness of 25 KPa, which is about twice higher than the paused time 3.15 ± 0.69 s (M ± SD) for the membrane stiffness of 2.5 KPa (t-test, p < 0.0008, **Figure 8**). The total deviation of all the nodes in the deformed shape in Figure 8a is 3.5 μm compared with the reference configuration, and in Figure 8b is 0.28 μm. Thus, these simulation results indicated that softer cells have prolonged average paused time.

*4.3.2 Effect of the number of GPIbα receptors expressed on the platelet membrane*

The interaction between platelet glycoprotein (GP) Ib-IX-V complex and vWF is the first step of the hemostatic response to vessel injury. As resting platelets interact with vWF, binding of vWF to GPIbα initiate platelet activation [40]. Meanwhile, in platelet-type von Willebrand disease, mutations of GPIb functional receptors can compromise hemostasis by increasing the affinity for vWF [41-42]. Some studies demonstrated that abnormalities in the concentrations of GPIb membrane proteins are present in patients with myeloproliferative disorders. In particular, decreased GPIb concentrations were found in patients with thrombocythaemia and leukemia [43-44]. How the platelet-substrate adhesion dynamics and subsequent platelet activation are affected by the number of GPIb is not clear. The objective of our simulations performed in this section was to gain insight into this problem. We varied the platelet receptor number from 10688 (normal) to 5344 (insufficient), and performed simulations with 30 different random seeds to obtain 30 different paused times under flow shear stress of 3.0 dyn/cm$^2$. The results of our simulations revealed that the paused time in the case of decreased receptor number was 2.07 ± 0.41s (M ± SD), which was significantly lower than 3.15 ± 0.69 s (M ± SD) for normal receptor number group (t-test, p < 0.02, **Figure 9**). Our simulations predicted that as the number of GPIb on the platelet membrane decreased, the paused time of platelet adhesion to vessel wall also decreased. Thus, the results of our model suggest that the number of functional GPIb is an important factor determining platelet adhesion and subsequent activation. This has important biological consequences, as controlling the number of functional GPIb receptors can provide means for development of novel anti-thrombotic drugs. The mechanism of these drugs is based on inhibiting/promoting the function of platelet GPIb receptors to decrease/increase adhesion of platelets to vWF to control blood clot growth [45].

*4.3.3 Effect of the platelet-platelet adhesion*

To study how platelet-platelet interaction affects platelet adhesion to the blood vessel wall we modeled dynamics of two platelets near the surface of the vessel (**Figure 10a**). In the model, the two platelets interacted with each other and one of them adhered to the vessel wall. Our simulations revealed that the platelet paused time was 1.61 ± 0.46 s (M ± SD) in the case of two adhesive platelets, which was significantly lower than the pause time of 3.15 ± 0.69 s (M ± SD) calculated for a single platelet interacted with the wall (t-test, p < 0.02, **Figure 10b**). These results indicate an important mechanism by which a single platelet adhesion can be affected due to interaction with neighboring cells. These findings have direct biological consequences and help to explain how the increased platelet concentration in blood can affect platelet-wall adherence.

## 5. Discussion

This paper describes a novel 3D model coupling processes at three biologically important spatial scales critical for early blood clot development and uses model to provide predictive simulations. First, our model provides a comprehensive representation of mechanical properties of a platelet based on the implementation of a hybrid membrane submodel to describe mechanical behavior of the cytoskeleton network and the lipid bilayer of the platelet. In previous studies, platelets were modeled as rigid bodies [22, 28, 46]. However, it has been experimentally shown [9] that platelets exhibited both elastic and viscoelastic behavior and that they underwent large deformation in shear flow [10].

Experimental studies demonstrated [1, 50-51] that flow shear stress could increase both bond formation and dissociation rates during platelet adhesion to the vessel wall. Additionally, estimates for the forces acting on platelet-substrate bonds were provided in [8]. However, [8] did not describe a detailed computational model to simulate the binding dynamics under various flow conditions. By combining 3D multiscale model with microfluidic experiments we provided a methodology to quantify in detail single platelet flipping in blood flow and platelet tethering to the injured vessel wall. It results in a two-way coupled fluid-cell interaction submodel combined with a stochastic submodel of formation/breakage of individual receptor-ligand bonds. This approach provided a biologically justified description of a platelet dynamics, which can be used to simulate dynamics of platelets under more complex flow conditions.

By incorporating physiological parameter values characterizing cellular membrane mechanics our method provides explicit representation for the structure of the cytoskeleton and simulation of cellular dynamics. Thus, our model allows one to examine how the mobility of cells is affected by their membrane structural and mechanical properties and hence, aids in providing prognostic assessment in blood cells disorders outcome. The model developed in this paper can be also used for simulating important biomedical problems which involve description of dynamics and deformation of cells in fluid flow including (patho)physiological inflammation involving leukocyte and platelet tethering to the vessel wall. Other important applications of the model can include studying cell aggregate formation in blood, metastasis of tumor cells as well as stem cell attachment to the target tissues.

**Acknowledgements:** This research was partially supported by the National Science Foundation grants DMS-0800612, DMS-1115887 and by the NIH grant 1R01GM095959. Oleg Kim was also partially supported by the College of Science, University of Notre Dame.

**Appendix: GPU implementation**

The CPU used for our simulation is Intel Xeon CPU L5520 with clock rate 2.27GHz. Our NVIDIA graphics card is GeForce GTX 480, Clock rate: 1.45GHz, CUDA Driver Version: 5.50, CUDA Runtime Version: 5.50, CUDA Capability version: 2.0. GPUs are separate devices with their own processors and memory devices which do not have direct access to the CPUs or CPUs' memory units. The communication pathway for transferring data between CPU memory and GPU memory has a relatively slow bandwidth capability compared to direct access to memory devices. Thus, it is necessary to minimize the communication as much as possible. The typical GPU code is composed of three main parts: 1) initialization, 2) execution and 3) cleanup. During initialization, model data is firstly allocated and initialized in CPU memory. CPU code then initializes connection to GPU device and allocates GPU memory for the model simulation data. The model data is copied from CPU memory to GPU memory units. During execution, GPU kernel functions are called and occasionally copy data between CPU and GPU memory devices. When a simulation is finished, both CPU and GPU memory units are freed and connection to GPU device is shut down for the cleanup.

**Figure 1** shows the flow chart of our simulation algorithm. For a single step of the simulation, its execution on GPU starts with hybrid membrane model. Assuming that a platelet consists of N triangle mesh elements and P nodes (each node represents a subcellular element (SCE)) (**Figure 2b**), The GPU kernel function to calculate forces in Eqs. (3) has the form:

*sem_Force_kernel<<< blocksPerGrid, threadsPerBlock >>>(grids->devImage){*

   *//Calculation of forces acting on subcellular elements*

   *... ...*

*}*

where *blocksPerGrid* and *threadsPerBlock* are determined according to the block and thread distribution on the GPU card, and P = *blocksPerGrid * threadsPerBlock*. If this is implemented on a single CPU in serial configuration, there will be total of P iterations for one step of simulation. In our GPU implementation, this simulation is performed simultaneously on P GPU threads. Hence it reduces the complexity of execution time from O(P) to O(1) for one step of simulation. Similarly, we use following GPU functions to calculate forces due to bending, area constraint and volume constraint:

*sem_Bending_kernel<<< blocksPerGrid, threadsPerBlock >>>(grids->devImage){*

   *//Calculation of forces due to bending*

   *... ...*

*}*

*sem_Area_Volume_kernel<<< blocksPerGrid, threadsPerBlock >>>(grids->devImage){*

   *//Calculation of forces due to area and volume constraint*

   ... ...

*}*

where *blocksPerGrid* and *threadsPerBlock* are determined according to the block and thread distribution on the GPU card, and N = *blocksPerGrid * threadsPerBlock*. It reduces the complexity of execution time from O(N) to O(1) comparing with CPU code.

    The forces acting on solid nodes are spread to its neighbor fluid nodes using Immersed Boundary method. The GPU kernel function has the form:

*fluid3d_force_distribute_kernel<<<blocksPerGrid3D,threadsPerBlock3D>>>(grids->devImage){*

    *//Implementation of IBM*

    ... ...

*}*

where both *blocksPerGrid3D* and *threadsPerBlock3D* have three dimensional structure similar to a 3D space coordinate. Let $blocksPerGrid3D = (x_b, y_b, z_b)$ and $threadsPerBlock3D = (x_t, y_t, z_t)$, and the fluid lattice has the size of $X \times Y \times Z$, then $X = x_b \times x_t$, $Y = y_b \times y_t$, and Z= $z_b \times z_t$. It reduces the complexity of execution time from O(XYZ) to O(1) comparing with CPU code. Similar strategy is applied to Lattice Boltzmann method implementation.

    There are M receptors in each of the N triangle mesh elements of a platelet. The GPU kernel function for dynamic adhesion model has the form:

*sem_platelet_wall_kernel<<<blocksPerGrid, threadsPerBlock>>>(grids->devImage) {*

     *//Implementation of DAM for a single receptor*

     ... ...

*}*

where $N \times M = blocksPerGrid \times threadsPerBlock$. It can speed up the execution time NM times comparing with CPU implementation.

    In conclusion, the GPU will reduce the whole simulation from $O(P) + O(N) + O(XYZ) + O(NM) \sim O(XYZ)$ to $O(1)$ in time complexity. **Table 3** shows real execution time of 10,000 steps simulation on both CPU and GPU for three different fluid grid sizes. As Table 3 showed, the execution time of GPU code is only about 1/100 of the CPU version for the fluid grid size we used in this paper. While CPU execution time increased linearly with fluid grid size, the GPU execution time increased much slower than CPU.     .

**Table 1.** Biological processes and submodels at different scales.

| Scales | Processes | Submodels | Coupling |
|---|---|---|---|
| < 0.1 µm<br>Sub-Cellular Level<br>Nanoscale | Ligand-Receptor<br>Interactions | Stochastic dynamic<br>adhesion model | 1. Nano-Micro scales:<br>coupled by explicitly<br>modeling receptors on<br>platelet membrane<br>nodes. |
| ~1 µm<br>Cellular Level<br>Microscale | Individual platelet<br>deforming, flipping and<br>adhering to vessel wall | Hybrid membrane<br>model | 2. Micro-Macro scales:<br>coupled through<br>Immersed Boundary |
| > 10 µm<br>Macroscale | Blood flow and its<br>interaction with platelet | Lattice Boltzmann<br>method | method |

**Table 2.** Values of physical parameters used in simulations

| Parameters | Definition | Value | References |
|---|---|---|---|
| a | Platelet radius | 1.0 μm | [21] |
| λ | Platelet aspect ratio | 0.25 | [20] |
| Υ | Flow shear rate | 300 and 400 s$^{-1}$ | [1] |
| ρ | Blood plasma density | 1.0239 g/cm$^3$ | [22] |
| μ | Blood plasma viscosity | 1.2 cP | [22] |
| E | Platelet elastic modulus | 25 kPa | [9] |
| $l_0$ | Average length of initial spring length | 75 nm | |
| $k_s$ | Global area constraint coefficient | $6000\,\dfrac{k_B T}{l_0^2}$ | [23] |
| $k_t$ | Local area constraint coefficient | $6000\,\dfrac{k_B T}{l_0^2}$ | [23] |
| $k_v$ | Volume constraint coefficient | $6000\,\dfrac{k_B T}{l_0^3}$ | [23] |
| $k_0$ | Bending modulus | $200 k_B T$ | [15] |
| T | Temperature | 300 K | |
| $k_{on}^0$ | Intrinsic cross-linking formation rate | $10^{-5}$ s$^{-1}$ | [1] |
| $k_{off}^0$ | Unstressed disassociation rate | 3.45 s$^{-1}$ | [1] |

**Table 3.** Execution time of 10,000 simulation steps on CPU and GPU for different fluid grid sizes

| Fluid grid size | CPU (s) | GPU (s) |
|:---:|:---:|:---:|
| $80 \times 300 \times 20$ | 36251 | 268 |
| $40 \times 150 \times 10$ | 2393 | 231 |
| $20 \times 75 \times 5$ | 232 | 51 |

**FIGURES**

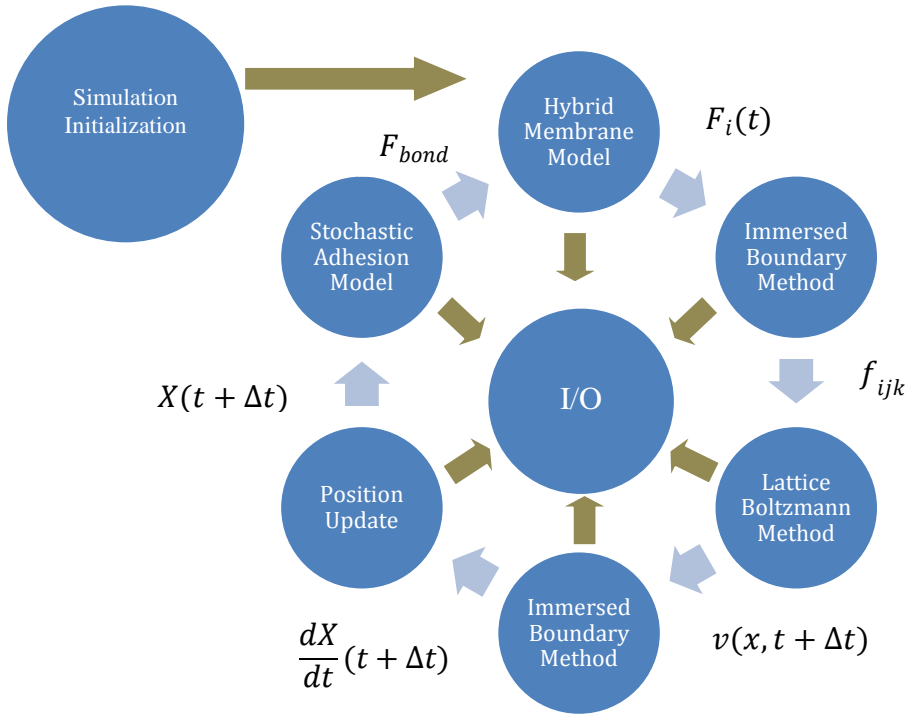

**Figure 1.** Flow chart of the simulation algorithm. Using hybrid membrane model, the forces $F_i(t)$ acting on cell elements was calculated. The forces $f_{ijk}$ acting on fluid node were spread from $F_i(t)$ by immersed boundary method. The velocity field $v(x, t+\Delta t)$ of fluid was obtained by Lattice Boltzmann method. The velocities of cell elements $\frac{dX}{dt}(t + \Delta t)$ were determined by immersed boundary interpolation. The cell elements positions were updated based on the velocities. Finally, stochastic adhesion model was used to determine the force $F_{bond}$ acting on receptor-ligand bond that binding platelet to vessel wall.

a)

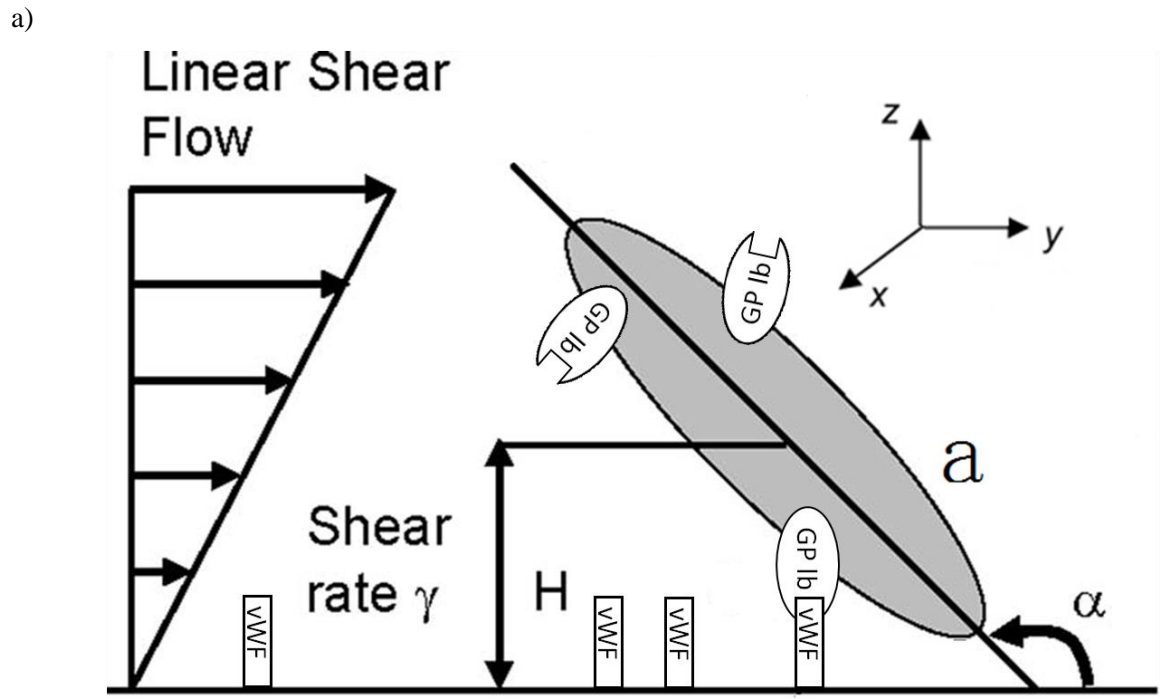

b)

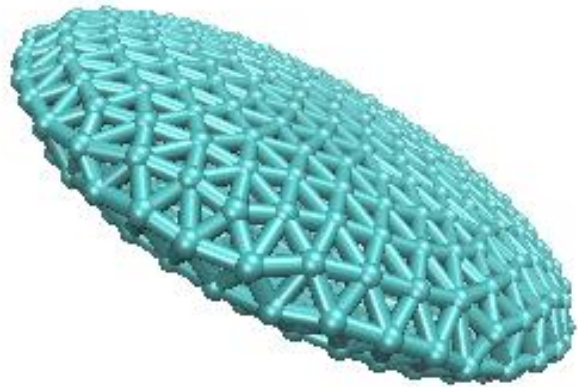

**Figure 2.** (a) Schematic diagram showing one platelet translating and rotating in shear flow near an infinite plane wall. (b) Structure of a platelet consisting of 958 of SCEs**.** The major radius, a, and centroid height, H, are defined as is the coordinate system and flow direction. One platelet is represented by a collection of elastically linked SCEs, interacting with one another via spring-like elastic force. The GPIbα receptors are randomly uniform distributed on platelet membrane, and vWF ligands are distributed on the wall.

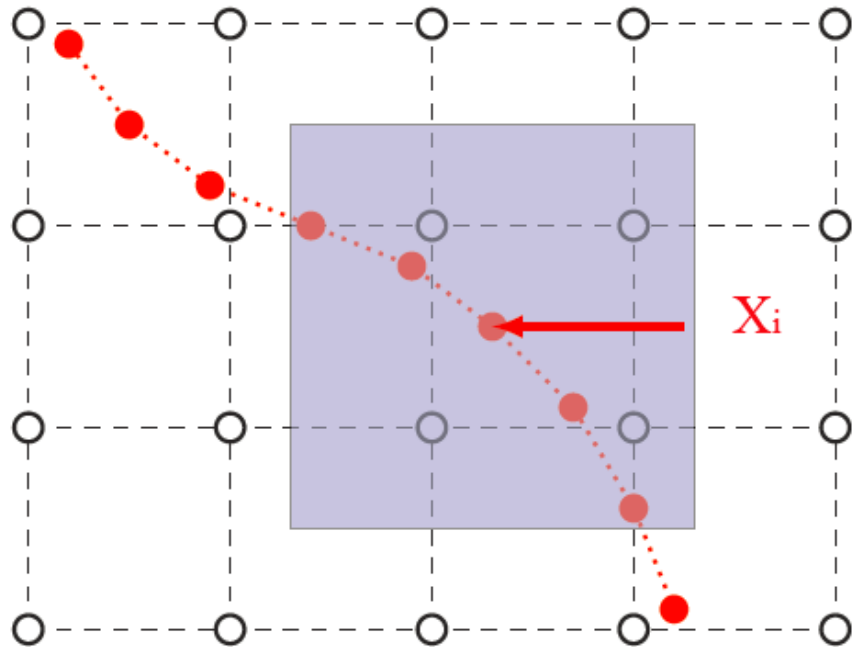

**Figure 3.** Eulerian fluid grid (black) and Lagrange solid elements (red). An Eulerian description is used for the fluid dynamics, and a Lagrangian description is used for objects immersed in the fluid. The communication between these two coordinate systems is realized by immersed boundary method.

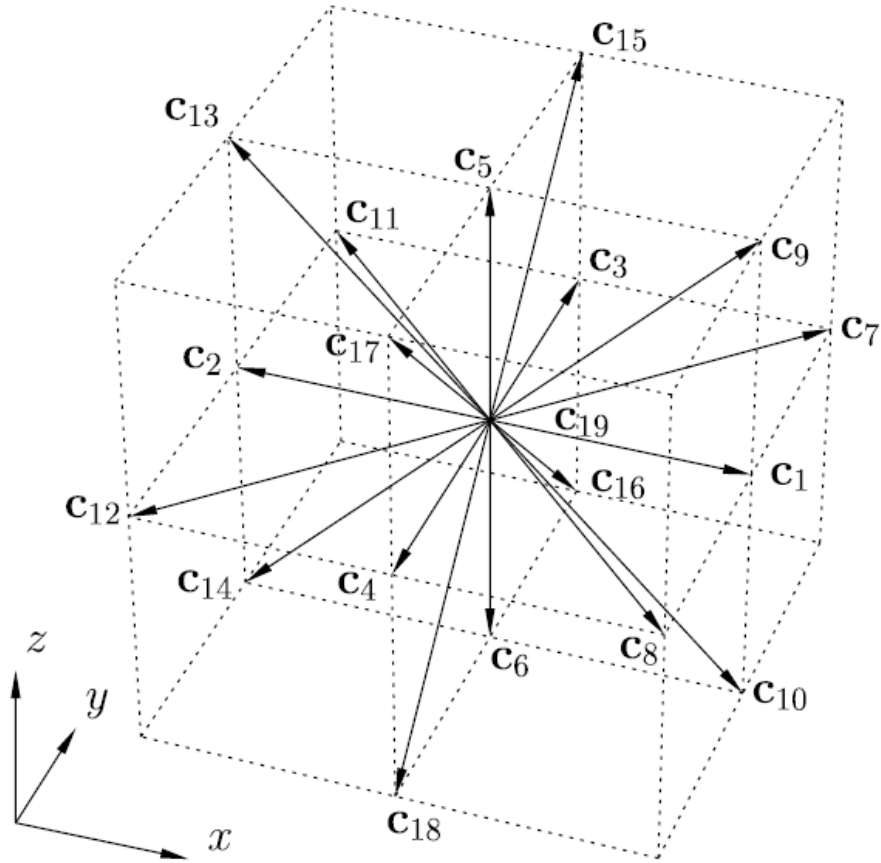

**Figure 4.** Lattice Boltzmann D3Q19 (3D and 19 velocities) model. The lattice vectors $c_i$ represent the velocities of the particles moving from the center grid point to its neighbor grid point.

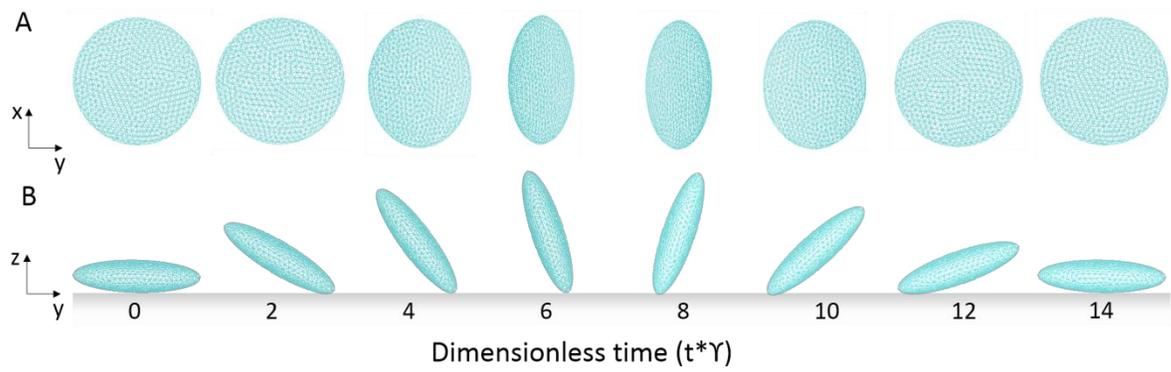

**Figure 5.** The configuration of the simulated platelets at different times flipping over the vessel wall for the wall shear stress of 3.0 dyn/cm². Image sequence A shows projection of the platelet on the x-y plane from dimensionless time point 0 to 14. Image sequence B shows projection of the platelet on the y-z plane from dimensionless time point 0 to 14. The coordinate system is defined in **Figure 2a**.

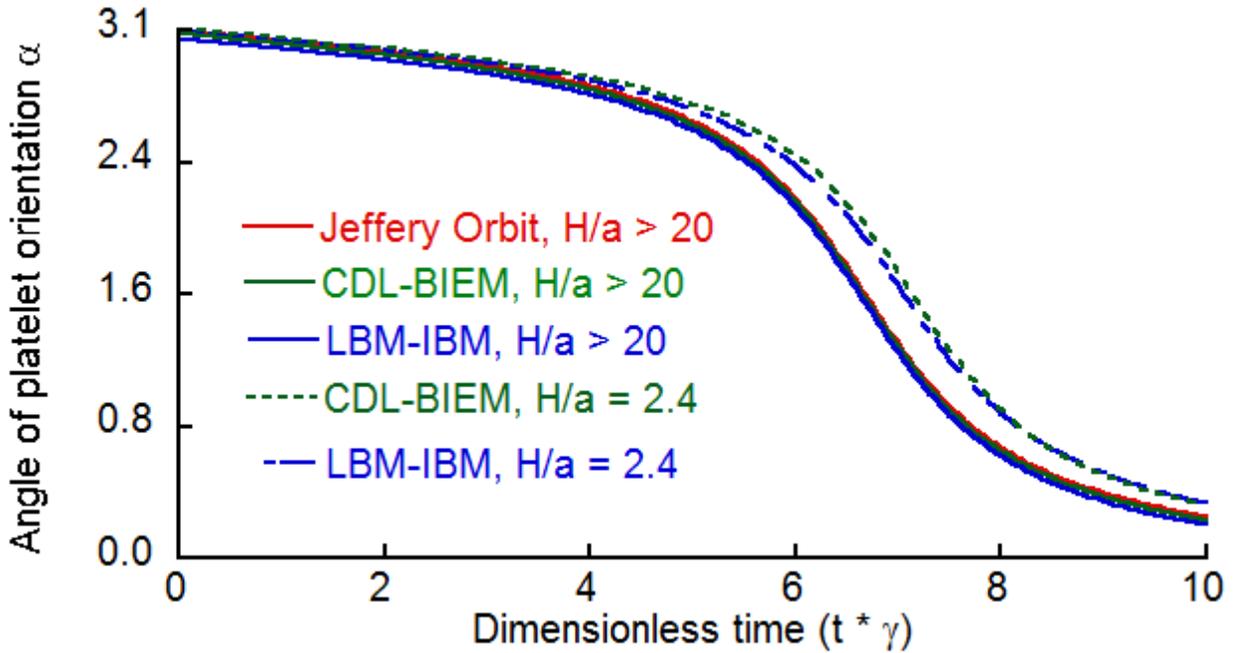

**Figure 6.** Validation of the platelet dynamics model. The analytical solution for the platelet rotational trajectory (Jeffery Orbit), trajectory calculated by a completed double layer–boundary integral equation method (CDL-BIEM) and our simulation (LBM-IBM) are shown by solid and dashed color lines (inset key).

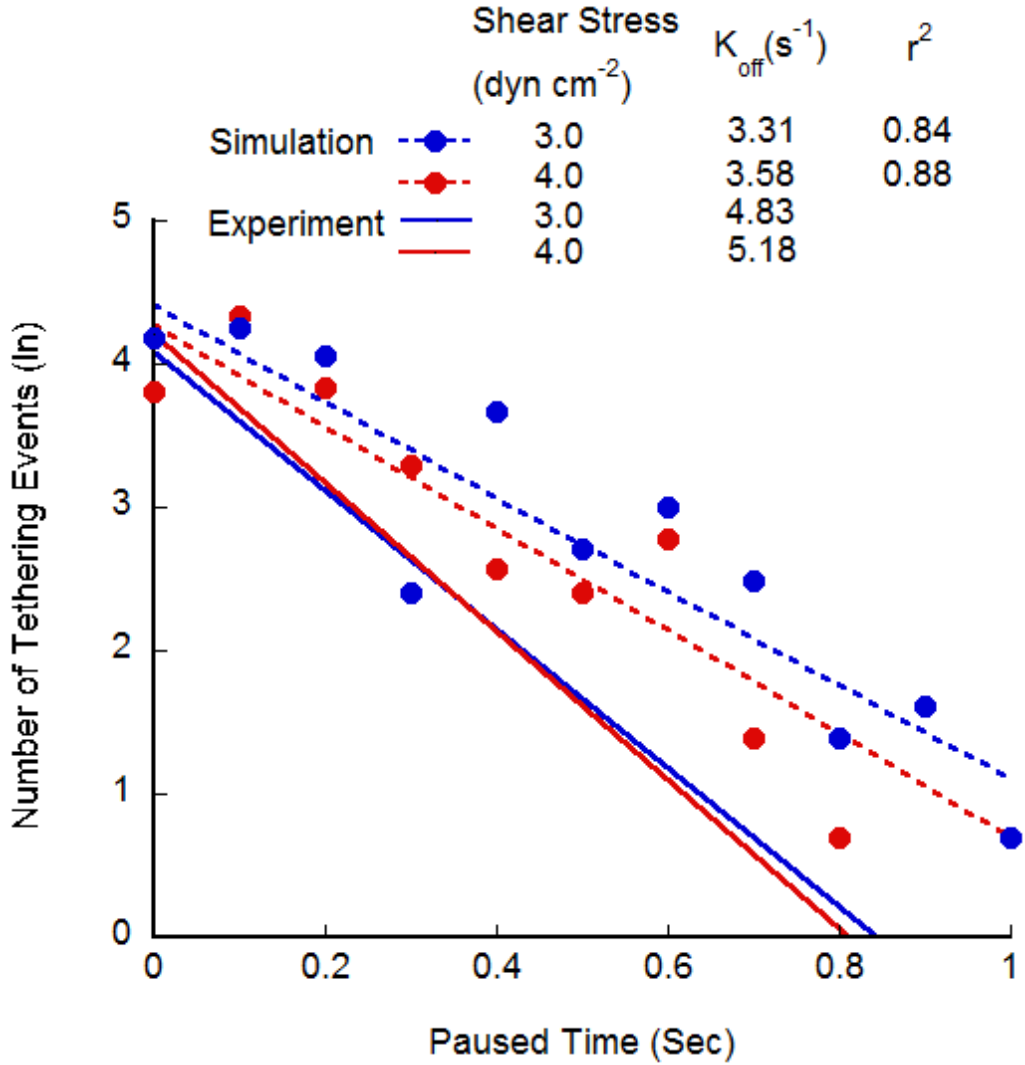

**Figure 7.** The number of tethering events as a function of the platelet paused time. The solid lines are the fitting lines of experimental data for shear stresses of 3.0 dyn cm$^{-2}$ (shown in blue) and 4.0 dyn cm$^{-2}$ (shown in red). The corresponding slopes of the fits ($k_{off}$ values) are -4.83 and -5.18. The dashed lines are the fitting lines of simulation results (shown with circles) for shear stresses of 3.0 dyn cm$^{-2}$ (shown in blue) and 4.0 dyn cm$^{-2}$. The corresponding slopes of the fits ($k_{off}$ values) are -3.31 and -3.58.

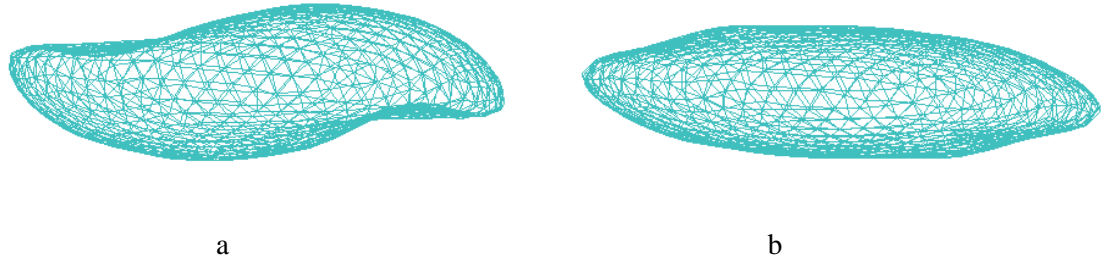

a                                                   b

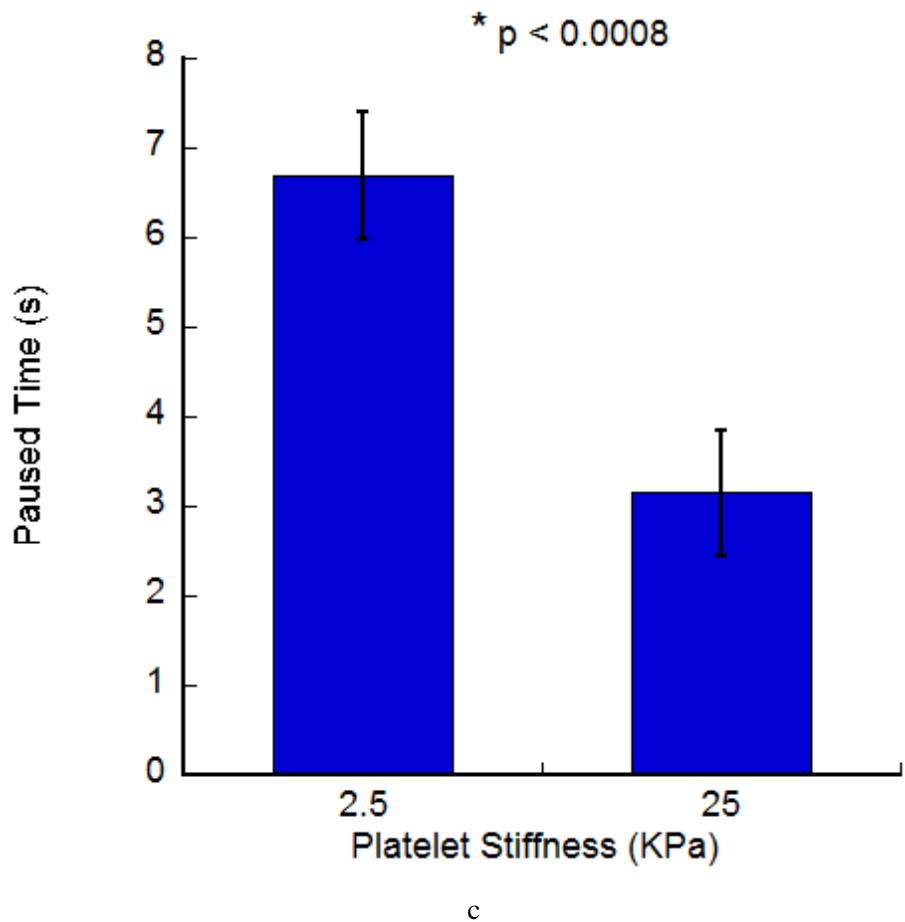

c

**Figure 8.** The simulated deformations of platelet structures during their adhesion to the vessel wall for platelet stiffness of 2.5 KPa (a) and 25 KPa (b).The effect of the platelet membrane stiffness on the platelet paused time (c). The paused time was $6.69 \pm 0.71$s (M $\pm$ SD) for the membrane stiffness of 25 KPa, which was about twice higher than the paused time of $3.15 \pm 0.69$ s (M $\pm$ SD) for the membrane stiffness of 2.5 KPa.

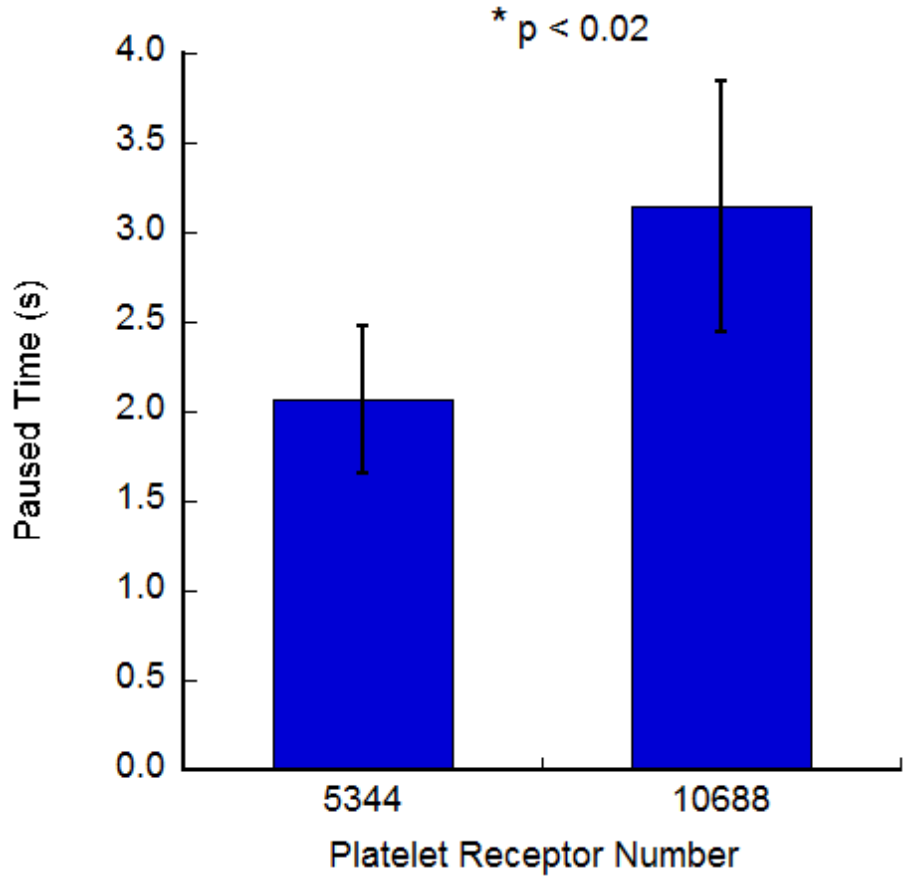

**Figure 9.** The effect of the number of platelet receptors on the platelet-vessel wall paused time. The platelet paused time for a decreased number of GPIb functional receptors was 2.07 ± 0.41s (M ± SD), which was significantly lower than the paused time of platelets having the normal number of receptors (3.15 ± 0.69 s, M ± SD).

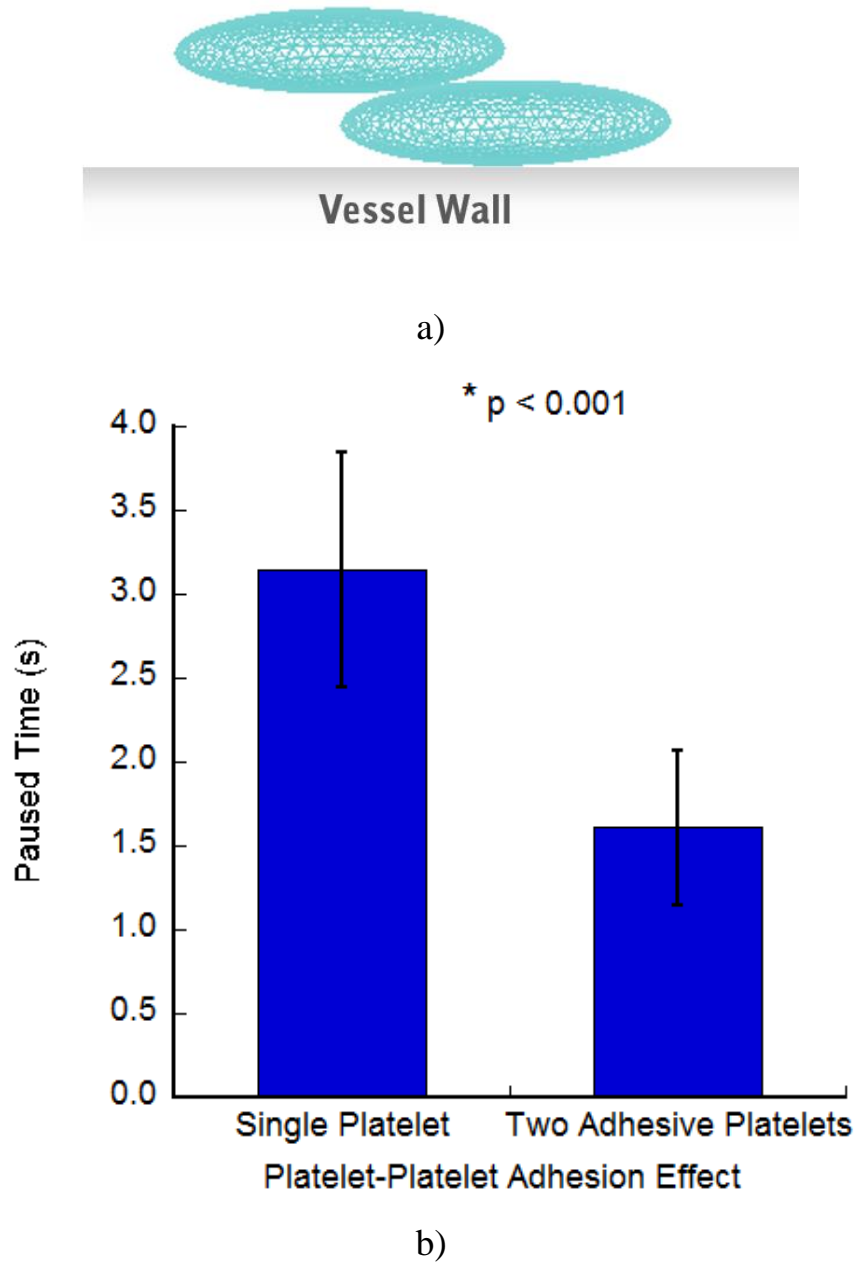

a)

b)

**Figure 10.** a) Initial configuration of two platelets used in simulations studying the effect of mutual interaction of platelets on platelet-wall adhesion. b) Platelet-vessel wall paused time as a function of the number of interacting platelets. The platelet paused time was $1.61 \pm 0.46$ s (M $\pm$ SD) for two adhesive platelets, which was significantly lower than the stopping time of platelets having the normal number of receptors ($3.15 \pm 0.69$ s, M $\pm$ SD).

**References**

[1] Doggett, T.A., Girdhar, G., Lawshe, A., Schmidtke, D.W., Laurenzi, I.J., Diamond, S.L. & Diacovo, T.G. 2002 Selectin-like kinetics and biomechanics promote rapid platelet adhesion in flow: the GPIb(alpha)-vWF tether bond. *Biophys J* **83**, 194-205.

[2] Hussain, M.A., Agnihotri, A. & Siedlecki, C.A. 2005 AFM imaging of ligand binding to platelet integrin alphaIIbbeta3 receptors reconstituted into planar lipid bilayers. *Langmuir* **21**, 6979-6986.

[3] Holland, N.B., Siedlecki, C.A. & Marchant, R.E. 1999 Intermolecular force mapping of platelet surfaces on collagen substrata. *J Biomed Mater Res* **45**, 167-174.

[4] Vinckier, A. & Semenza, G. 1998 Measuring elasticity of biological materials by atomic force microscopy. *FEBS Lett* **430**, 12-16.

[5] Yago, T., Lou, J., Wu, T., Yang, J., Miner, J.J., Coburn, L., Lopez, J.A., Cruz, M.A., Dong, J.F., McIntire, L.V., et al. 2008 Platelet glycoprotein Ibalpha forms catch bonds with human WT vWF but not with type 2B von Willebrand disease vWF. *J Clin Invest* **118**, 3195-3207.

[6] Xu, Z., Kim, O., Kamocka, M., Rosen, E.D. & Alber, M. 2012 Multiscale models of thrombogenesis. *Wiley Interdiscip Rev Syst Biol Med* **4**, 237-246.

[7] Xu, Z., Kamocka, M., Alber, M. & Rosen, E.D. 2011 Computational approaches to studying thrombus development. *Arterioscler Thromb Vasc Biol* **31**, 500-505.

[8] Mody, N.A., Lomakin, O., Doggett, T.A., Diacovo, T.G. & King, M.R. 2005 Mechanics of transient platelet adhesion to von Willebrand factor under flow. *Biophys J* **88**, 1432-1443.

[9] Radmacher, M., Fritz, M., Kacher, C.M., Cleveland, J.P. & Hansma, P.K. 1996 Measuring the viscoelastic properties of human platelets with the atomic force microscope. *Biophys J* **70**, 556-567.

[10] Lettinga, M.P., Holmqvist, P., Ballesta, P., Rogers, S., Kleshchanok, D. & Struth, B. 2012

Nonlinear behavior of nematic platelet dispersions in shear flow. *Phys Rev Lett* **109**, 246001.

[11] Sandersius, S.A. & Newman, T.J. 2008 Modeling cell rheology with the Subcellular Element Model. *Phys Biol* **5**, 015002.

[12] Sweet, C.R., Chatterjee, S., Xu, Z., Bisordi, K., Rosen, E.D. & Alber, M. 2011 Modelling platelet-blood flow interaction using the subcellular element Langevin method. *J R Soc Interface* **8**, 1760-1771.

[13] Xu, Z., Christley, S., Lioi, J., Kim, O., Harvey, C., Sun, W., Rosen, E.D. & Alber, M. 2012 Multiscale model of fibrin accumulation on the blood clot surface and platelet dynamics. *Methods Cell Biol* **110**, 367-388.

[14] Hao, W., Xu, Z., Lin, G. & Liu, C. A Fictitious Domain Method with a Hybrid Cell Model for Simulating Motion of Cells in Fluid Flow. *In preparation*.

[15] Du, Q., Liu, C., Ryham, R. & Wang, X.Q. 2009 Energetic variational approaches in modeling vesicle and fluid interactions. *Physica D* **238**, 923-930.

[16] King, M.R. & Hammer, D.A. 2001 Multiparticle adhesive dynamics. Interactions between stably rolling cells. *Biophys J* **81**, 799-813.

[17] Dembo, M., Torney, D.C., Saxman, K. & Hammer, D. 1988 The reaction-limited kinetics of membrane-to-surface adhesion and detachment. *Proc R Soc Lond B Biol Sci* **234**, 55-83.

[18] Yago, T., Wu, J., Wey, C.D., Klopocki, A.G., Zhu, C. & McEver, R.P. 2004 Catch bonds govern adhesion through L-selectin at threshold shear. *J Cell Biol* **166**, 913-923.

[19] Savage, B., Saldivar, E. & Ruggeri, Z.M. 1996 Initiation of platelet adhesion by arrest onto fibrinogen or translocation on von Willebrand factor. *Cell* **84**, 289-297.

[20] Frojmovic, M., Longmire, K. & van de Ven, T.G. 1990 Long-range interactions in mammalian platelet aggregation. II. The role of platelet pseudopod number and length. *Biophys J*

**58**, 309-318.

[21] Popel, A.S. & Johnson, P.C. 2005 Microcirculation and Hemorheology. *Annu Rev Fluid Mech* **37**, 43-69.

[22] Mody, N.A. & King, M.R. 2005 Three-dimensional simulations of a platelet-shaped spheroid near a wall in shear flow. *Phys Fluids* **17**.

[23] Li, J., Dao, M., Lim, C.T. & Suresh, S. 2005 Spectrin-level modeling of the cytoskeleton and optical tweezers stretching of the erythrocyte. *Biophys J* **88**, 3707-3719.

[24] Wu, J.S. & Aidun, C.K. 2010 Simulating 3D deformable particle suspensions using lattice Boltzmann method with discrete external boundary force. *Int J Numer Meth Fl* **62**, 765-783.

[25] Le, D.V., White, J., Peraire, J., Lim, K.M. & Khoo, B.C. 2009 An implicit immersed boundary method for three-dimensional fluid-membrane interactions. *J Comput Phys* **228**, 8427-8445.

[26] Reininger, A.J., Heijnen, H.F., Schumann, H., Specht, H.M., Schramm, W. & Ruggeri, Z.M. 2006 Mechanism of platelet adhesion to von Willebrand factor and microparticle formation under high shear stress. *Blood* **107**, 3537-3545.

[27] Berndt, M.C., Shen, Y., Dopheide, S.M., Gardiner, E.E. & Andrews, R.K. 2001 The vascular biology of the glycoprotein Ib-IX-V complex. *Thromb Haemost* **86**, 178-188.

[28] Mody, N.A. & King, M.R. 2008 Platelet adhesive dynamics. Part II: high shear-induced transient aggregation via GPIbalpha-vWF-GPIbalpha bridging. *Biophys J* **95**, 2556-2574.

[29] Singh, I., Shankaran, H., Beauharnois, M.E., Xiao, Z., Alexandridis, P. & Neelamegham, S. 2006 Solution structure of human von Willebrand factor studied using small angle neutron scattering. *J Biol Chem* **281**, 38266-38275.

[30] Fox, J.E., Aggerbeck, L.P. & Berndt, M.C. 1988 Structure of the glycoprotein Ib.IX

complex from platelet membranes. *J Biol Chem* **263**, 4882-4890.

[31] Hammer, D.A. & Apte, S.M. 1992 Simulation of cell rolling and adhesion on surfaces in shear flow: general results and analysis of selectin-mediated neutrophil adhesion. *Biophys J* **63**, 35-57.

[32] Guo, Z., Zheng, C. & Shi, B. 2002 Discrete lattice effects on the forcing term in the lattice Boltzmann method. *Phys Rev E Stat Nonlin Soft Matter Phys* **65**, 046308.

[33] Feng, Y.T., Han, K. & Owen, D.R.J. 2007 Coupled lattice Boltzmann method and discrete element modelling of particle transport in turbulent fluid flows: Computational issues. *Int J Numer Meth Eng* **72**, 1111-1134.

[34] Hecht, M. & Harting, J. 2010 Implementation of on-site velocity boundary conditions for D3Q19 lattice Boltzmann simulations. *J. Stat. Mech.*

[35] Peskin, C.S. 1972 Flow Patterns around Heart Valves - Numerical Method. *J Comput Phys* **10**, 252-&.

[36] Peskin, C.S. 2002 The immersed boundary method. *Acta Numerica* **11**, 479-517.

[37] Interlandi, G. & Thomas, W. 2010 The catch bond mechanism between von Willebrand factor and platelet surface receptors investigated by molecular dynamics simulations. *Proteins* **78**, 2506-2522.

[38] Pereverzev, Y.V., Prezhdo, O.V., Forero, M., Sokurenko, E.V. & Thomas, W.E. 2005 The two-pathway model for the catch-slip transition in biological adhesion. *Biophys J* **89**, 1446-1454.

[39] Shamova, E.V., Gorudko, I.V., Drozd, E.S., Chizhik, S.A., Martinovich, G.G., Cherenkevich, S.N. & Timoshenko, A.V. 2011 Redox regulation of morphology, cell stiffness, and lectin-induced aggregation of human platelets. *Eur Biophys J* **40**, 195-208.

[40] Kroll, M.H., Harris, T.S., Moake, J.L., Handin, R.I. & Schafer, A.I. 1991 von Willebrand

factor binding to platelet GpIb initiates signals for platelet activation. *J Clin Invest* **88**, 1568-1573.

[41] Kumar, R.A., Dong, J.F., Thaggard, J.A., Cruz, M.A., Lopez, J.A. & McIntire, L.V. 2003 Kinetics of GPIbalpha-vWF-A1 tether bond under flow: effect of GPIbalpha mutations on the association and dissociation rates. *Biophys J* **85**, 4099-4109.

[42] Andrews, R.K., Lopez, J.A. & Berndt, M.C. 1997 Molecular mechanisms of platelet adhesion and activation. *Int J Biochem Cell Biol* **29**, 91-105.

[43] Mazzucato, M., De Marco, L., De Angelis, V., De Roia, D., Bizzaro, N. & Casonato, A. 1989 Platelet membrane abnormalities in myeloproliferative disorders: decrease in glycoproteins Ib and IIb/IIIa complex is associated with deficient receptor function. *Br J Haematol* **73**, 369-374.

[44] Jensen, M.K., de Nully Brown, P., Lund, B.V., Nielsen, O.J. & Hasselbalch, H.C. 2000 Increased platelet activation and abnormal membrane glycoprotein content and redistribution in myeloproliferative disorders. *Br J Haematol* **110**, 116-124.

[45] Ji, X. & Hou, M. 2011 Novel agents for anti-platelet therapy. *J Hematol Oncol* **4**, 44.

[46] Mody, N.A. & King, M.R. 2008 Platelet adhesive dynamics. Part I: characterization of platelet hydrodynamic collisions and wall effects. *Biophys J* **95**, 2539-2555.

[47] Fogelson, A.L. & Guy, R.D. 2004 Platelet-wall interactions in continuum models of platelet thrombosis: formulation and numerical solution. *Math Med Biol* **21**, 293-334.

[48] Sui, Y., Chew, Y.T. & Low, H.T. 2007 A lattice Boltzmann study on the large deformation of red blood cells in shear flow. *Int J Mod Phys C* **18**, 993-1011.

[49] Skorczewski, T., Erickson, L.C. & Fogelson, A.L. 2013 Platelet motion near a vessel wall or thrombus surface in two-dimensional whole blood simulations. *Biophys J* **104**, 1764-1772.

[50] Alevriadou, B.R., Moake, J.L., Turner, N.A., Ruggeri, Z.M., Folie, B.J., Phillips, M.D., Schreiber, A.B., Hrinda, M.E. & McIntire, L.V. 1993 Real-time analysis of shear-dependent thrombus formation and its blockade by inhibitors of von Willebrand factor binding to platelets. *Blood* **81**, 1263-1276.

[51] Kroll, M.H., Hellums, J.D., McIntire, L.V., Schafer, A.I. & Moake, J.L. 1996 Platelets and shear stress. *Blood* **88**, 1525-1541.